\documentclass[superscriptaddress,twocolumn,prl]{revtex4}
\usepackage{amsmath}
\usepackage{amssymb}
\usepackage{graphicx}
\usepackage{hyperref}
\usepackage{amsfonts}
\usepackage{bbm}
\usepackage{doi}

\begin{document}
\title{Contact Topology and the Classification of Disclination Lines in Cholesteric Liquid Crystals}
\author{Joseph Pollard}
\email{joseph.pollard@durham.ac.uk}
\affiliation{Department of Physics, Durham University, Durham, DH1 3LE, United Kingdom}
\author{Gareth P. Alexander}
\email{G.P.Alexander@warwick.ac.uk}
\affiliation{Department of Physics and Centre for Complexity Science, University of Warwick, Coventry, CV4 7AL, United Kingdom.}

\begin{abstract}
  We give a topological classification of defect lines in cholesteric liquid crystals using methods from contact topology. By focusing on the role played by the chirality of the material, we demonstrate a fundamental distinction between `tight' and `overtwisted' disclination lines not detected by standard homotopy theory arguments. The classification of overtwisted lines is the same as nematics, however, we show that tight disclinations possess a topological layer number that is conserved as long as the twist is nonvanishing. Finally, we observe that chirality frustrates the escape of removable defect lines, and explain how this frustration underlies the formation of several structures observed in experiments.
\end{abstract}
\date{\today}
\maketitle

Many liquid crystal textures are distinguished by topological invariants derived from homotopy theory~\cite{mermin1979,kurik1988,kleman1989}. The classification of defects and solitons, both two- and three-dimensional, using homotopy groups is central to the modern understanding of liquid crystals and their properties~\cite{alexander2012,machon2016PRSA,chen2013,wu2022}. However, despite the successes of homotopy theory methods, it has long been recognised that they are insufficient to fully describe materials with a spatially modulated ground state, such as smectics and cholesterics~\cite{poenaru1981,chen2009,machon2019}.

Cholesterics are characterised by the property that the director field ${\bf n}$ has everywhere a uniform sense of twist, ${\bf n} \cdot \nabla \times {\bf n} < 0$ for a right-handed material. This constraint on the twist is central to the rich morphology of structures displayed by cholesteric materials~\cite{wright1989,copar2011PRE,copar2011PRL,sec2012,darmon2016,darmon2016b,posnjak2017,tai2019,wu2022}, but it is not accounted for in homotopy theory arguments, and standard topological invariants often fail to distinguish between qualitatively distinct cholesteric configurations~\cite{machon2017}. The non-vanishing of the twist implies that the director defines a `contact structure'~\cite{geiges2008}, and the field of contact topology provides a set of techniques for analysing such directors that goes beyond homotopy theory. Techniques and insights from the field of contact topology are becoming increasingly important in the study of cholesteric materials: they have been used to demonstrate the preservation of the layer structure in a cholesteric~\cite{machon2017}; to describe chiral point defects and elucidate the role of boundary conditions in the stability of complex defect structures in spherical droplets~\cite{pollard2019}; to explain the stability of Skyrmions in liquid crystals and chiral magnets~\cite{hu2021}; to analyse defect structures in cylindrical capillaries~\cite{eun2021}; and to shed light on the transition pathways between different cholesteric textures~\cite{han2022}.

A classification of defect lines in cholesterics was first given by Kleman \& Friedel~\cite{kleman1969} who identified three classes, the $\chi$ lines (defects in the director), $\lambda$ lines (defects in the pitch axis), and $\tau$ lines (defects in the director and pitch), and subsequently placed within the homotopy theory of defects~\cite{volovik1977,bouligand1978}. This approach puts the pitch axis on equal footing with the director field, even though it is only the director that appears in the free energy. More subtly, it doesn't build in a consistent handedness (sense of twist) for the director. More recent geometric approaches take the pitch axis to be derived from the director gradients~\cite{efrati2014,beller2014,machon2016prx} and constrain the handedness by adopting methods of contact geometry~\cite{machon2017,pollard2019,hu2021,eun2021,han2022} and it is this approach that we employ here. We show that the local classification of disclinations in cholesterics splits into two distinct cases, corresponding to tight and overtwisted contact structures. For overtwisted disclinations, the classification is the same as in nematics. However, tight disclinations possess a topological layer number that is conserved as long as the twist does not vanish, resulting in a much finer topological classification.

Our classification can be contrasted with the nematic case, where there is no constraint on the handedness. In nematics there are four homotopy classes for the local structure on the tubular neighbourhood of a defect loop~\cite{janich1987,copar2011PRL,alexander2012}. Representatives for each class are
\begin{equation}
    {\bf n} = \cos \bigl( \tfrac{1}{2} \theta + \tfrac{\nu}{2} \phi \bigr) \,{\bf e}_x + \sin \bigl( \tfrac{1}{2} \theta + \tfrac{\nu}{2} \phi \bigr) \,{\bf e}_y ,
    \label{eq:dir_janich}
\end{equation}
where $\theta$, $\phi$ are the meridional and longitudinal angles, and $\nu \in \mathbb{Z}_4$ is the J\"anich index. In the case $\nu=0$ this extends to the global director field for a charge zero defect loop first given by Friedel \& de Gennes~\cite{friedel1969,binysh2020}. It is a feature of these that the form of the director field~\eqref{eq:dir_janich} is independent of the geometry of the defect loop, whose shape and orientation can be arbitrary relative to the $xy$-plane in which the director rotates. Such defect loops are not chiral and ${\bf n} \cdot \nabla \times {\bf n}$ takes both signs in the neighbourhood of the defect~\cite{binysh2020}. In contrast, chiral defect loops have a director structure that is more closely connected to the geometry of the defect line in order to maintain a consistent handedness.

\begin{figure}[t]
  \centering
  \includegraphics[width=0.95\linewidth]{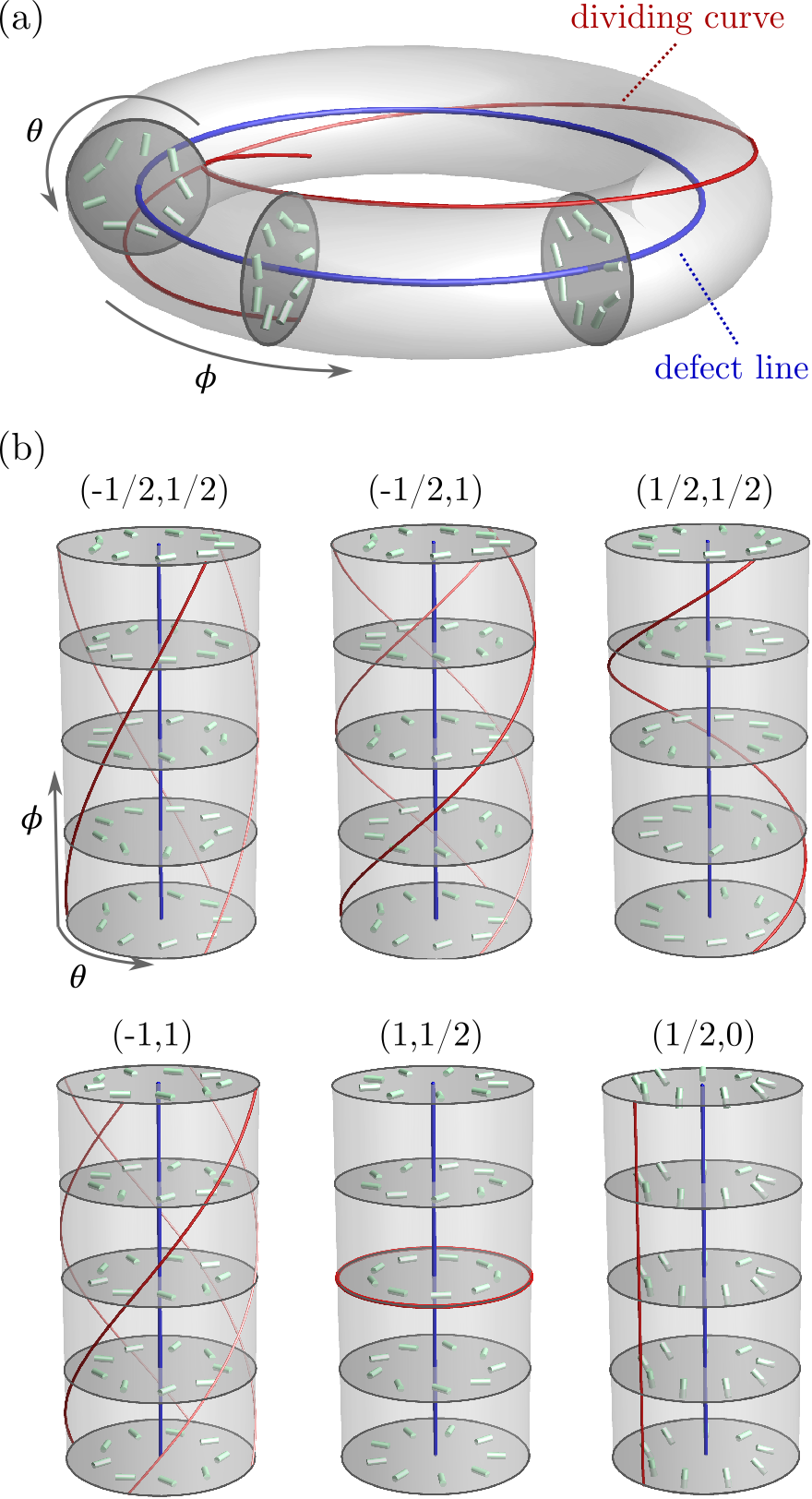}
  \caption{(a) Schematic of the tubular neighbourhood of a defect line (blue) with the director (green cylinders) shown on some cross-sections and part of the dividing curve (red). The schematic corresponds to the representative texture ${\bf n}_{-1/2,1}$. (b) Examples from the family of tight cholesteric disclinations~\eqref{eq:tight_model_k_q} for a selection of values of $(k,q)$, shown in abstracted, standardised form. The examples include the `exceptional' cases $k=1$ and $q=0$~\eqref{eq:tight_model_k_0}, where the dividing curve has either zero or infinite slope. Where the dividing curve has more than one component, one of them has been displayed in darker shade to aid visualisation.}
  \label{fig:disclination}
\end{figure}

A tubular neighbourhood of the defect loop is a solid torus (see the schematic in Fig.~\ref{fig:disclination}(a)) and so it suffices to classify textures on $D^2 \times S^1$, with a singular line along the central axis $0 \times S^1$, up to homotopies that fix the singular line and such that the twist ${\bf n} \cdot \nabla \times {\bf n}$ never vanishes. This classification splits into two cases: the `tight' case and the `overtwisted' case, reflecting a fundamental dichotomy in contact topology. The distinction between tight and overtwisted director fields is technical~\cite{machon2017} but informally a director is overtwisted if it contains a Skyrmion with a full $\pi$-twist. Contact topology tells us that tight director fields cannot be transformed into overtwisted ones without either the introduction of additional defects or the twist density ${\bf n} \cdot \nabla \times {\bf n}$ vanishing~\cite{machon2017}. Further, Eliashberg's theorem~\cite{eliashberg1989}, a classical result in contact topology, implies that overtwisted disclination lines have the same classification as nematic defects and do not possess any additional invariants as a result of their chirality. Thus the classification of overtwisted disclination loops is the same as that for nematics and given by a J\"anich index $\nu \in \mathbb{Z}_4$.
%WE ARE NOT GIVING (CHIRAL) REPRESENTATIVES FOR THESE CLASSES
%
It remains to classify the tight disclinations. These are given by distinct classes of dividing curves on the boundary of the tubular neighbourhood of the defect to be described presently; see also Fig.~\ref{fig:disclination}.

\begin{figure*}[t]
  \centering
  \includegraphics[width=\linewidth]{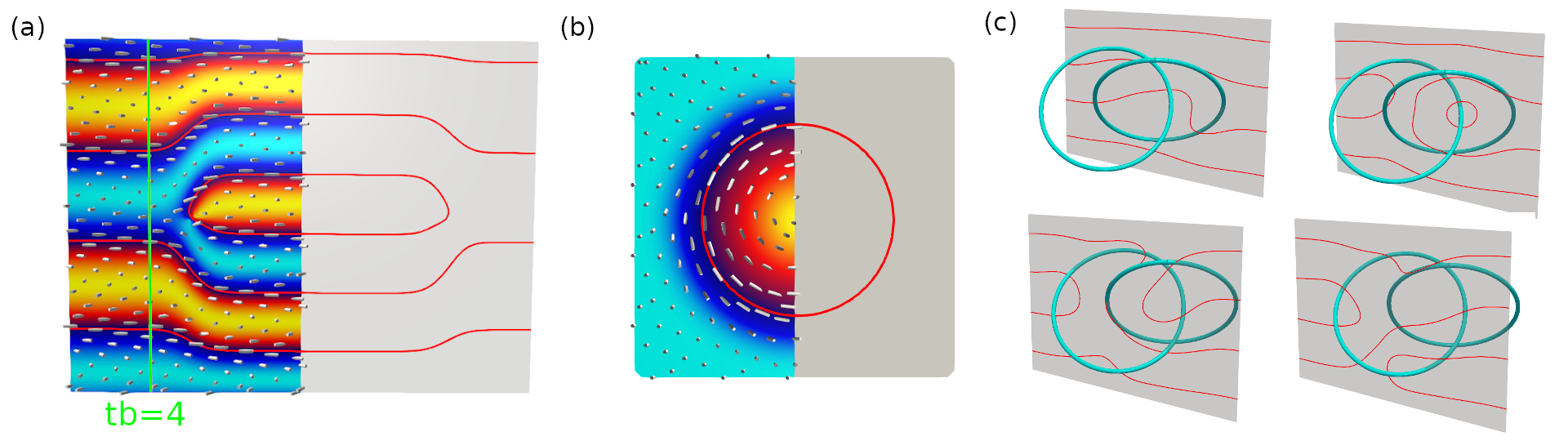}
  \caption{Illustrations of convex surfaces, their dividing curves, and the computation of the Thurston-Bennequin (TB) number. (a) A convex torus (grey) cutting across a cholesteric texture exhibiting a change in the number of layers. The director, shown as white sticks, points out from the surface in orange regions and into the surface in blue regions. The topological content of this data is entirely captured by the dividing curve (red), the set of points where the director is tangent to the convex surface. The green line indicates an example calculation of a Thurston-Bennequin (TB) number. (b) On a convex surface cutting across a Skyrmion, the dividing curve is a closed contractible loops. By Giroux's criterion the contact structure is overtwisted. (c) Convex surface tomography of a Hopfion. As the convex surface is slid across the Hopfion, changes in the dividing curve track the changes in the layer structure. The presence of a closed component bounding a disk (top right) reveals the presence of a Skyrmion, with a local structure equivalent to that shown in panel (b) and shows that the Hopfion is overtwisted. The pale blue curves indicate the linked $\lambda^{+1}$ lines of the Hopfion.}
  \label{fig:convex}
\end{figure*}

We first describe representatives of each of the homotopy classes and then describe how the classification is obtained using contact topology. We use polar coordinates $(r,\theta)$ for points of $D^2$ and an angular coordinate $\phi$ along $S^1$. We take the solid angle framing~\cite{binysh2018} to define the zero of the local azimuthal angle $\theta$ and let $\{{\bf e}_1, {\bf e}_2, {\bf e}_3\}$ be an adapted orthonormal frame with ${\bf e}_3$ tangent to the defect line and ${\bf e}_1$ the normal vector in the direction $\theta=0$.
The representatives can be distinguished according to the direction of the pitch axis, either parallel or perpendicular to the defect line, the former being the case for $\chi$ lines and the latter for $\tau$ lines. When the pitch is parallel to the defect axis we have a two-parameter family (of $\chi$ lines)
\begin{equation}
 {\bf n}^{\chi}_{k, q} = \cos(k\theta + q \phi) \,{\bf e}_1 + \sin(k\theta + q \phi) \,{\bf e}_2 ,
\label{eq:tight_model_k_q}
\end{equation}
where $k$ is an arbitrary half-integer, the winding number of the director in the disk $D^2$, and $q>0$ is a positive half-integer giving the number of twists of the director along the $S^1$. When the pitch is perpendicular to the defect line, representatives can be given as the one-parameter family
\begin{equation}
    {\bf n}^{\textrm{az}}_{k} = \cos(k\theta) \,{\bf e}_3 + \sin(k\theta) \,{\bf e}_{r} ,
\label{eq:tight_model_k_0}
\end{equation}
where the pitch axis is oriented azimuthally. Here, $k>0$ is a positive half-integer and ${\bf e}_r$ is the unit radial vector normal to the defect line. The more familiar form of $\tau$ lines is given by instead taking the pitch axis along a vector field ${\bf m}_k$ in the $\{{\bf e}_1,{\bf e}_2\}$-plane with winding number $k$ (any half-integer) and setting
\begin{equation}
    {\bf n}^{\tau}_{k} = \sin\psi \,{\bf e}_3 + \cos\psi \,{\bf m}_{k} \times {\bf e}_3 ,
    \label{eq:tight_model_tau}
\end{equation}
where $\psi$ is a helical phase increasing along the pitch axis.
Several examples are shown in Fig.~\ref{fig:disclination}(b).
%For the first family~\eqref{eq:tight_model_k_q} the pitch axis is in the tangential direction ${\bf e}_3$, transverse to the cholesteric `layers', and the disclination has a `screw-like' structure. For the second family~\eqref{eq:tight_model_k_0} the pitch axis is in the local azimuthal direction ${\bf e}_{\theta}$ and the disclination is `edge-like'.

Examples arise naturally in a Grandjean-Cano wedge geometry~\cite{bouligand1974,masuda1995,smalyukh2002}, where the cholesteric ground state is disrupted by the introduction of defects. These may be a $\lambda^{-1/2} \lambda^{+1/2}$ pair, but can also be pairs of defects consisting of a $\lambda^{\pm 1/2}$ line and a disclination line of type ${\bf n}^{\tau}_{\mp 1/2}$. Patterned substrates can be used to stabilise webs of disclinations of type ${\bf n}^{\tau}_{+ 1/2}$~\cite{wang2017}, and disclinations of type ${\bf n}^{\tau}_{-1/2}$ occur around colloidal inclusions, in the well-known saturn ring texture~\cite{copar2011PRE,copar2011PRL}. Here, the homeotropic anchoring on the boundary of colloid implies the existence of a region of revered-handedness close to the colloid~\cite{pollard2019}, which region may either be nonsingular, or coupled to the defect itself. Disclinations with the same structure as the type ${\bf n}^\chi_{-1/2, 1}$ disclination have been generated around colloidal inclusions~\cite{copar2011PRE,copar2011PRL}, and ${\bf n}^\chi_{-1/2, 1}$ disclinations also occur in all of the known blue phases~\cite{wright1989}.

We now give a proof of this classification result. The proof makes use of three concepts from contact topology: convex surface theory, the Thurston-Bennequin invariant and the tight-overtwisted dichotomy. We describe these concepts briefly; a more complete account can be found in~\cite{geiges2008}. Convex surface theory is a general tool for studying topological and geometrical properties of ordered media~\cite{giroux1991,geiges2008}. Consider an embedded surface $S$, either closed or with boundary orthgonal to the director, that that does not intersect any defects. The director will be tangent to $S$ along a collection of disjoint curves $\Gamma$ that divide $S$ into regions where the director points out of the surface, a set $S^+$, and regions where it points into the surface, $S^-$. This situation is generic: $S$ is called a convex surface and $\Gamma$ a dividing curve.
A convex surface cutting across a cholesteric texture containing several $\lambda$-lines is shown in Fig.~\ref{fig:convex}(a). The director is shown in the left half of the panel, with colours indicating whether it points into (blue) or out of (orange) the surface; on the right half of the panel we show only the dividing curve (red).

The dividing curve allows us to compute local topological information. For instance, it determines the Skyrmion charge $Q$ on $S$ via the formula
\begin{equation}
 2Q = \chi(S^+) - \chi(S^-).
\label{eq:skyrmion_charge}
\end{equation}
This is illustrated for a single Skyrmion in Fig.~\ref{fig:convex}(b). The region $S^+$ is a disk, with $\chi(S^+) =1$. The remainder is a punctured plane, with $\chi(S^-) = -1$, so that~\eqref{eq:skyrmion_charge} gives $Q=1$. A similar calculation shows that the director in Fig.~\ref{fig:convex}(a) also has $Q=1$. In the case of a spherical surface the same formula~\eqref{eq:skyrmion_charge} determines the point defect charge enclosed by $S$~\cite{copar2012}.

These properties of convex surfaces and dividing curves apply generally, including for achiral materials. However, in chiral materials we can also compute invariants connected to the non-zero twist. One is the Thurston-Bennequin invariant, which informally represents a count of the number of cholesteric layers~\cite{machon2017}. Given a closed curve $C$ that is everywhere orthogonal to the director, called a Legendrian curve, its Thurston-Bennequin number $\textrm{tb}(C)$ is the number of right-handed $\pi$ rotations of the director as one moves around $C$~\cite{tb_note}. For such a curve on a surface which intersects the dividing curve transversely its Thurston-Bennequin number is the count of intersections between the curves
\begin{equation}
    \textrm{tb}(C) = \bigl| \Gamma \cap C \bigr| .
\label{eq:tb}
\end{equation}
An example computation is shown in Fig.~\ref{fig:convex}(a) for a standard cholesteric texture with $\lambda$ lines and an extra layer. The minimum value of $\textrm{tb}(C)$ over all curves in a given isotopy class, which we denote $\overline{\textrm{tb}}$, is a topological invariant~\cite{geiges2008,machon2017}. For the example shown, this is a count of the number of cholesteric layers.

More complex textures, such as the Hopfion shown in Fig.~\ref{fig:convex}(c), can be visualised by studying the dividing curves on a series of slices through the material, a process called `tomography'. Changes in the number of components of the dividing curve, and thus the minimal Thurston-Bennequin number attainable for a curve on each given slice, reveal fundamental changes in local topology.

\begin{figure}[t]
  \centering
  \includegraphics[width=\linewidth]{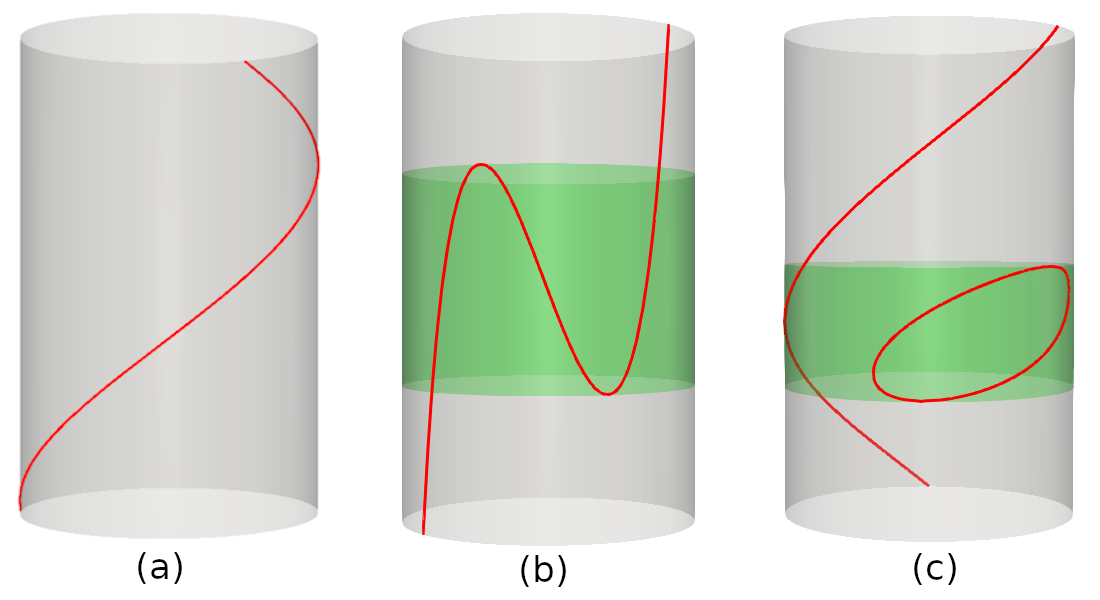}
  \caption{Three motifs occur in the dividing curve (red) on a section of a convex torus. (a) The dividing curve turns everywhere counter-clockwise and the winding $k_\phi$ is constant. (b) The dividing curve turns back on itself, indicating a region (green) in which $k_\phi =-1/2$, while $k_\phi = +1/2$ in the remainder of the panel. (c) There is a separate component of the dividing curve which bounds a disk on the torus, also indicating the presence of a region (green) with $k_\phi =-1/2$.}
  \label{fig:dividing_curve}
\end{figure}

For a disclination, the boundary of a tubular neighbourhood serves as a convex surface $S$. Generically, the director will not be orthogonal to the disks $D_{\phi}$ of constant $\phi$ and hence we can take its projection into these disks. We refer to this projection as the profile of the line. On each $D_{\phi}$ the profile winds around the disclination with some half-integer winding number $k_{\phi}$. As the dividing curve is identified with the points where the director is tangent to $S$, the number of intersections between $\Gamma$ and the boundary of $D_{\phi}$ is $2|1-k_{\phi}|$. The winding $k_\phi$ need not be constant and can vary with $\phi$. We distinguish two cases: when $k_{\phi}$ is the same for all $\phi$, and when it varies. In the latter case there are two ways in which the number of intersections can change: from a `kink', as in Fig.~\ref{fig:dividing_curve}(b); or from a separate component of the dividing curve that bounds a closed disk, as in Fig.~\ref{fig:dividing_curve}(c). A result of Honda~\cite{honda2000} establishes that `kinks' can always be removed by a homotopy, so that this case reduces to that of constant $k_{\phi}$.

The second case, where a component of $\Gamma$ bounds a disk, is more fundamental and indicates that the disclination line is overtwisted. This result is known as Giroux's criterion~\cite{giroux1991,geiges2008}, which states that a component of the dividing curve bounding a disk appears if and only if there is an overtwisted disk close to the convex surface. %: Technically, it is possible to find a closed Legendrian curve $C$, bounding a disk, with $\textrm{tb}(C) = 0$.
The director on the disk bounded by $\Gamma$ is a Skyrmion tube, with profile equivalent to that in Fig.~\ref{fig:convex}(b), that terminates on the disclination line.
%For a further example, this criterion demonstrates that each director shown in Fig.~\ref{fig:convex} is overtwisted.
A fundamental result of Eliashberg~\cite{eliashberg1989} states that overtwisted directors do not have any additional contact topological invariants and hence their classification is the same as that of nematics: they are classified locally by a J\"anich index $\nu \in \mathbb{Z}_4$~\cite{note_point_defects,pollard_thesis}. Thus, overtwistedness distinguishes singular lines with an essentially-varying $k_\phi$ from those for which $k_\phi$ is constant or can be made constant by the removal of kinks. In the latter case, the director is tight in a neighbourhood of the singular line.

For the tight case, the dividing curve is isotopic to a rational line on the torus, or a set of rational lines all with the same slope. The examples~\eqref{eq:tight_model_k_q},~\eqref{eq:tight_model_k_0} cover all such possibilities. We will show that they have different Thurston-Bennequin invariants; it then follows from standard results of contact topology~\cite{geiges2008} that they are not homotopic. For a longitudinal Legendrian curve $C$ on $S$ with zero linking number with the disclination, its minimal Thurston-Bennequin number is $\overline{\textrm{tb}} = 2q$ for the screw-like ($\chi$) defects~\eqref{eq:tight_model_k_q} and zero for the edge-like ($\tau$) defects~\eqref{eq:tight_model_k_0}. For a meridional Legendrian curve, the minimal Thurston-Bennequin number is $\overline{\textrm{tb}} = 2|1-k|$ for both screw-like and edge-like defects. Since these invariants take different values for different pairs of $k,q$ (including the edge-type as $q=0$), the different models ${\bf n}^{\chi}_{k,q}$ and ${\bf n}^{\textrm{az}}_{k}$ all represent homotopically distinct chiral directors.

%To demonstrate that ${\bf n}^\chi_{k,q}$ for different $k,q$ are not homotopic, we consider Legendrian curves $C$ on $S$ and their Thurston-Bennequin numbers given by~\eqref{eq:tb}. First consider a longitude with zero linking number with the disclination. We readily see that the minimum Thurston-Bennequin number for any such curve is $\overline{\textrm{tb}} = 2q$.
%We readily see that the minimum Thurston-Bennequin number for any such curve is $\overline{\textrm{tb}} = 4q|1-k|$.
%By standard results of contact topology~\cite{geiges2008}, it then follows that ${\bf n}^\chi_{k,q}$, ${\bf n}^\chi_{k,q^\prime}$ are not equivalent if $q\neq q^\prime$.
%Next, consider a meridional curve $C$. This time $\overline{\textrm{tb}} = 2|1-k|$, which establishes that ${\bf n}^\chi_{k,q}$, ${\bf n}^\chi_{k^\prime,q}$ are not equivalent if $k\neq k^\prime$.

%{\bf NEEDS TO BE CHANGED}

%The pair ${\bf n}_{\pm k, 0}$ is an exception to the above formulas. For both directors the minimal Thurston-Bennequin number for a longitudinal curve is 0, and for a meridional curve it is 1, implying these directors are homotopic; the family $\cos t \ {\bf n}_{+k,0} + \sin t \ {\bf n}_{-k,0}$ with $t \in [0,\pi/2]$ furnishes a homotopy between them.

This completes the topological classification of singular lines in a cholesteric. We conclude with a discussion of the consequences of this result beyond just the classification of chiral disclinations.

Our classification covers both non-orientable (half-integer $k$) and orientable (integer $k$) singular lines. For integer lines in a nematic the singularity in the director can be removed by `escape in the third dimension'. This process extends to cholesterics only for negative windings; when $k$ is positive the chiral escape is frustrated~\cite{pollard_thesis}. The frustration arises when the the projection of a director into a surface has a purely radial singularity---per results from contact topology, this cannot occur in a nonsingular chiral director~\cite{geiges2008,eun2021}. Consequently, the removal of such a singular line either results in regions where the twist has the wrong handedness~\cite{pollard2019}, or else requires the introduction of point defects; the latter is energetically preferred; for a more extended discussion see~\cite{pollard_thesis}.

%Taking $k$ to be an integer in the family ${\bf n}^\chi_{k,q}$ produces a family of orientable singular lines, and our classification applies to this case as well as the case of $k = \pm 1/2$. Such lines are not stable defects in a nematic material, as they can be removed by an `escape into the third dimension'. This process is frustrated in a cholesteric when both $k$ and $q$ are positive. The frustration arises when the the projection of a director into a surface has a purely radial singularity---per results from contact topology, this cannot occur in a nonsingular chiral director~\cite{geiges2008,eun2021}. Consequently, the removal of such a singular line either results in regions where the twist density is positive, or else requires the introduction of point defects; the latter is energetically preferred.

A singular line of type ${\bf n}^{\tau}_{1}$ may be removed by escape, and the resulting texture is the core of a double twist cylinder. For $q > 0$, a ${\bf n}^\chi_{+1,q}$ line is replaced by a string of $2q$ point defects of alternating charge. When $q=1$, the resulting pair of defects is a toron~\cite{smalyukh2010}. Longer strings of point defects arising from the removal of this type of singularity occur in cholesteric droplets and shells~\cite{sec2012,darmon2016}, where the spherical geometry naturally promotes the formation of a ${\bf n}^\chi_{+1,q}$ line, and have recently been observed in chromonic liquid crystals in a cylindrical geometry~\cite{eun2021}.

Orientable singularities of winding $+2$ also arise naturally in a spherical geometry. A ${\bf n}^\chi_{+2,q}$ singularity will pull itself apart into a pair of $+1$ singular lines of type ${\bf n}^{\tau}_{1}$ which are linked $q$ times, in a process analogous to the conversion of twist into writhe for a space curve. These ${\bf n}^{\tau}_{1}$ lines may then escape to form a pair of linked double twist tubes. In a spherical droplet, the resulting texture is known as the spherulitic or Frank-Pryce texture~\cite{robinson1956,sec2012,darmon2016,darmon2016b}. The knotted structure in the recently-observed heliknotons~\cite{tai2019,wu2022} also results from this process, with the linking of the two meron tubes giving rise to a nonzero Hopf invariant. Another possibility, in which the $+2$ singular lines splits into a pair of disclinations with windings $+3/2$ and $+1/2$, has also been observed~\cite{darmon2016}.

Similar considerations would govern the removal singular lines of winding $+3, +4, $ etc---although no suitable geometry exists to stabilise such structures---as well as degenerate disclinations of type ${\bf n}^\chi_{+3/2,q}$, whose decomposition produces a generic disclination linked with a $\lambda$ line. These structures have been observed in spherical shells~\cite{darmon2016}.

Finally, the dichotomy between overtwisted and tight disclinations has implications for the crossing of disclination lines. Classical homotopy theory arguments applied to cholesterics show that crossing two disclinations produces a $\lambda^{+1}$ line tether connecting them~\cite{poenaru1977}. Consequently, if we take two tight disclination lines in a cholesteric and pass one through the other, we end up with a pair of disclination lines tethered to a $\lambda^{+1}$ line, which implies the disclination lines are overtwisted after the crossing event. Further analysis of the crossing and reconnection of defects using methods of contact topology would be an important extension of Ref.~\cite{poenaru1977}.

% Due to the complexity of describing defect crossing in materials with nonabelian fundamental group it is not immediately apparent whether this process always results in the creation of a region of reversed handedness, as would accompany the creation of $\lambda^{+1}$ lines in a defect-free tight cholesteric state; for instance, it is possible to nucleate a pair of point defects in a tight background without creating regions of reversed handedness.

%GPA: I AM CUTTING THIS FOR NOW ...
%We have given a topological classification of disclination lines in cholesteric materials based on the assumption of non-zero twist distortion, extending the previous classification based on homotopy theory. Our classification uses methods from contact topology which have a broad applicability to the study of chiral systems in general. The classification reveals a fundamental distinction between tight and overtwisted disclinations---the later possess a chiral topological invariant, which we further relate to the formation of strings of point defects and helical structures in spherical droplets and shells.

%The director in a neighbourhood of a disclination in a cholesteric is tightly coupled to the geometry of the disclination line---this is not the case for an achiral nematic. A full appreciation of this fact remains for future work. The homotopy theory classification implies that the crossing of two tight disclinations results in overtwisted disclinations, now tethered by a $\lambda$ line---the contact topology perspective on such crossing events remains to be explored.

\acknowledgments{This work was supported by the UK EPSRC through Grant No. EP/L015374/1. JP supported by a Warwick IAS Early Career Fellowship.}

\end{document}